\documentclass[12pt]{article}
\usepackage{graphicx}
\newcommand{\be}{\begin{equation}}
\newcommand{\ee}{\end{equation}}
\newcommand{\ba}{\begin{eqnarray}}
\newcommand{\ea}{\end{eqnarray}}
\newcommand{\ed}{\end{document}}

\newcommand{\bfr}{\begin{flushright}}
\newcommand{\efr}{\end{flushright}}
\newcommand{\bfl}{\begin{flushleft}}
\newcommand{\efl}{\end{flushleft}}
\renewcommand{\baselinestretch}{1}
\date{}

\begin{document}
\title{Soliton solutions of nonlinear Schr\"odinger equation on simple networks}
\author{Z.Sobirov, D. Matrasulov, K.Sabirov\\
Heat Physics Department, Uzbek Academy of Sciences\\
28 Katartal Street,100135 Tashkent,Uzbekistan\\
S. Sawada\\
Department of Physics, Kwansei Gakuin University\\ Sanda 669-1337, Japan\\
 K.Nakamura\\
Department of Applied Physics, Osaka City University\\ Osaka
558-8585, Japan}

\maketitle

\begin{abstract}
We show soliton solutions of nonlinear Schr\"odinger equation on
simple networks consisting of vertices and bonds, where the
strength of cubic nonlinearity is different from bond to bond. We
concentrate on reflectionless propagation of Zakharov-Shabat's
solitons through a branched chain, namely, a primary star graph
consisting of three semi-infinite bonds connected at a vertex. The
conservation of the norm and the global current elucidates: (1)
the solution on each bond is a part of the universal soliton
solution on a simple 1-dimensional (1-d) chain but multiplied by
the inverse of square root of bond-dependent nonlinearity; (2)
nonlinearities at individual bonds around each vertex must satisfy
a sum rule.  Under these conditions, all other conservation rules
for a simple 1-d chain have proved to hold for multi-soliton
solutions on graphs. The argument is extended to other graphs,
i.e., general star graphs, tree graphs, loop graphs and their
combinations. Numerical evidence is also given on the
reflectionless propagation of a soliton through a branched
chain.\end{abstract}

\section{Introduction}
Transport in networks with vertices and bonds \cite{har,kott}
received a growing attention recently. The practical importance of
this problem is caused by the fact that those networks mimic
networks of nonlinear waveguides and and optical fibers
\cite{kiv}, Bose-Einstein condensates in optical lattices
\cite{tro}, superconducting ladders of Josephson junctions
\cite{binder,bir}, double helix of DNA \cite{yak}, etc.

Most studies so far, however, are restricted to solving the linear
Schr\"odinger  equation to obtain the energy spectra in closed
networks and transmission probabilities for open networks with
semi-infinite leads.

On the other hand, with  introduction of the nonlinearity to the
time-dependent Schr\"odinger equation, the network provides a nice
playground where one can see interesting soliton propagations and
nonlinear dynamics in general. There already exist an accumulation
of numerical studies of the soliton propagation through the
discrete chain attached with small graphs
\cite{fla,abl,bur,pik,fla2}, where the discrete nonlinear
Schr\"odinger equation(DNLSE)   plays a role. However, we see
little exact analytical treatment of soliton propagation through
networks, namely an assembly of continuum line segments connected
at vertices, within a framework of nonlinear Schr\"odinger
equation(NLSE)\cite{sulem}. The subject is difficult due to the
presence of vertices where the underlying chain should bifurcate
or multi-furcate in general. A set of  the continuity and
smoothness conditions, which was exploited for  the linear
Schr\"odinger equation on graphs (see for example, Smilansky et
al.\cite{kott}), would provide a natural boundary condition at
each vertex. Then a soliton coming  into the vertex along one of
the bonds shows a complicated motion around the vertex.  When time
elapses, however,  one can expect stable solitons with smaller
amplitude propagating along each of bonds with one of them
representing the reflected soliton along the incoming bond. The
stable soliton in each bond is expected to be Zakharov-Shabat's
solitons (ZSSs) discovered in the nineteen-seventies\cite{zakh}.
Hereafter the terminology of ZSS will cover from a single soliton
through multi-soliton solutions presented in \cite{zakh}.
\begin{figure}
    \centerline{\includegraphics[width=6cm, height=25mm]{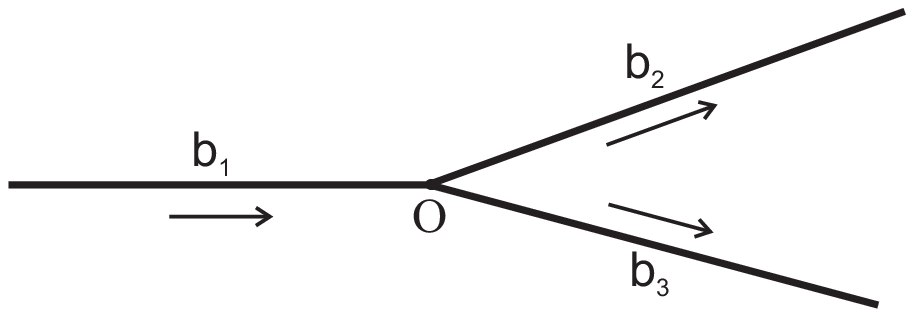} \ \ \ \ \includegraphics[width=6cm, height=30mm]{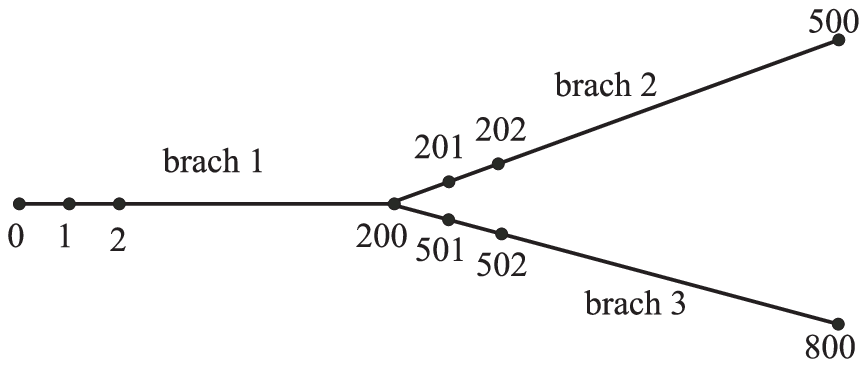}}
    \caption{Primary star graph. Left panel: 3 semi-infinite chains
    connected at a vertex $O$;
    Right panel: space-discrete version of the left panel being used
    for numerical simulations in Appendix.}
   \label{star}
\end{figure}

\begin{figure}
\centerline{\includegraphics[width=11 cm]{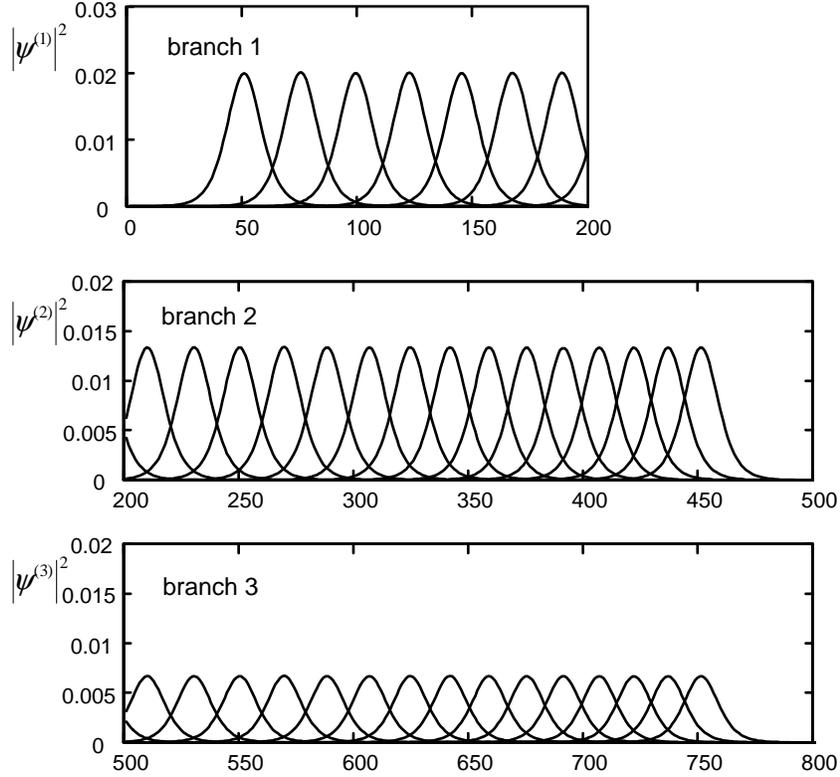}}
\caption{Time evolution of a soliton propagation through a vertex
(numerical result): an example. Space distribution of wave
function probability is depicted in every time interval $T=50.0$
with time used commonly in branches 2 and 3. Abscissa represents
discrete lattice coordinates defined in Fig.\ref{star}. Strength
of nonlinearity at each bond are $\beta_1=1,\beta_2=1.5,
\beta_3=3$. Initial profile is Zakharov-Shabat soliton in
Eq.(\ref{1-soliton}) at $t=0$ with parameters $a=0.1,v=0.1$. Time
difference in numerical iteration is $\Delta t=0.1$. For the
numerical method to solve nonlinear Schr\"odinger equation on
primary star graph, see Appendix.} \label{numerical}
\end{figure}

Interestingly,  under an appropriate  relationship among values of
nonlinearity at individual bonds together with a suitable boundary
condition at the vertex, one can see nonlinear dynamics of
solitons with no reflection at the vertex. For example, let us
consider an elementary branched chain (see Fig.\ref{star}),
namely, a primary star graph (PSG) consisting of three
semi-infinite bonds connected at the vertex $O$.

As shown in Fig.\ref{numerical}, our numerical simulation of DNLSE
on the discrete version of PSG indicates: the soliton starting at
lattice point $x=50$ in the branch 1 enters the vertex  at $x=200$
and is smoothly split into a pair of smaller solitons in the
branches 2 and 3, with neither reflection nor emergence of
radiation at the vertex! Inspired by this discovery, we shall
explore conditions among solitonic parameters and the relationship
among strengths of nonlinearity at individual bonds in order to
see the reflectionless propagation of  solitons through networks
or graphs.

In this paper, we shall present the exact analytical treatment of
soliton dynamics  in networks,  by concentrating on the
reflectionless propagation of ZSSs through vertices in the
networks. We shall search for the conditions for parameters
characterizing the ZSSs and the sum rule for strengths of
nonlinearity in each bond to satisfy the conservation rules for
the norm and current during their propagation through  networks.
Network models we shall choose  are  star graphs, tree graphs,
loop graphs and their combinations.  We assume bonds of star
graphs and edge bonds of tree graphs as semi-infinite, while other
intermediate tree and loop bonds are taken finite.

In Section II , using PSG, we shall show a basic idea of the
soliton propagation along the branched chain. ZSS along the
incoming chain is shown to  bifurcate at the vertex, resulting in
a pair of ZSSs with each propagating along the outgoing bonds. In
Subsection II.A we establish a connection formula for the current
density at the vertex and find a relation among  parameters
characterizing the ZSSs.  In Subsection II.B, we shall address the
additional condition to guarantee the total current conservation
rule, finding the sum rule for strengths of nonlinearity at each
bond. In Subsection II.C, the boundary condition at the vertex is
elucidated. In Section III, we show that the  energy and all other
conservations for a simple 1-dimensional (1-d) chain to hold for
general solitons through PSG. In Section IV, we investigate the
bifurcation of two-soliton  and $N$-soliton solutions at the
vertex of a branched chain. In Section V, the cases of tree
graphs,  loop graphs and their combinations are investigated,
where the soliton solution, sum rule and an infinite-number of
conservation rules will be constructed  by generalizing the result
for PSG.  Summary and discussion are devoted to Section VI.
Appendix is devoted to the way of numerically solving the
corresponding DNLSE on PSG.

\section{Single Soliton Propagation on Primary Star Graph}
\subsection{Norm Conservation Rule and Connection Formula for Current Density}

We consider an elementary branched chain or PSG in the upper panel of Fig.\ref{star}, where the vertex site is now taken as origin $O$.
Space coordinates in individual bonds  are here defined as $b_1 \sim
(-\infty, 0)$, $b_2 \sim (0,+\infty)$ and $b_2 \sim (0,+\infty)$.
On each bond we have the  nonlinear Schr\"odinger equation (NLSE)
\begin{equation}\label{eq1}
i\frac{\partial \Psi_k}{\partial t}+\frac{\partial^2
\Psi_k}{\partial x_{k}^2}+\beta_k |\Psi_{k}|^2\Psi_{k}=0, \ \
k=1,2,3,
\end{equation}
with $x_k$ defined on $-\infty<x_{1}<0, \ 0<x_{2},  x_{3}<\infty$.
It should be noted that the strength of nonlinearity $\beta_k (>0)$ may be different among bonds. The solution in PSG satisfies the following conditions at infinity:
$\Psi_1(x_1)\to 0$ at $x_1\to -\infty$, $\Psi_k(x_k)\to 0$ at
$x_k\to \infty$ for $ k=2,3$.  One of the physically important conditions for the solution in PSG is the
norm conservation. The norm is defined as
\begin{equation}
N=\|\Psi\|^2=\int\limits_{-\infty}^0|\Psi_1(x,t)|^2dx+
\int\limits_0^{\infty}|\Psi_2(x,t)|^2dx+\int\limits_0^{\infty}|\Psi_3(x,t)|^2dx.
\label{norm}
\end{equation}

Let us find conditions for which the norm is conservative. For
this purpose we calculate its time-derivative:
\begin{equation}
\frac{d}{dt}N=\int\limits_{-\infty}^0\frac{\partial
|\Psi_1(x,t)|^2}{\partial t}dx+
\int\limits_0^{\infty}\frac{\partial |\Psi_2(x,t)|^2}{\partial
t}dx+
\int\limits_0^{\infty}\frac{\partial|\Psi_3(x,t)|^2}{\partial
t}dx. \label{normdel}
\end{equation}

From Eq. (\ref{eq1}) we have the continuity equation,
\begin{equation}
\frac{\partial\left|\Psi_k(x,t)\right|^2}{\partial t}=-2\frac{\partial}{\partial x}{\rm Im}
\left[\Psi_k^*(x,t)\frac{\partial\Psi_k(x,t)}{\partial x}\right].
\label{cont}
\end{equation}

Using Eq.(\ref{cont}) in Eq.(\ref{normdel}), we have
\begin{equation}
\frac{d}{dt}N=-j_1(0,t)+j_2(0,t)+j_3(0,t),
\label{curren-cons1}
\end{equation}
where $j_k (k=1,2,3)$ is the current density defined by
\begin{equation}
j_k(x,t)=2{\rm Im}
\left[\Psi_k^*(x,t)\frac{\partial\Psi_k(x,t)}{\partial x}\right].
\label{current}
\end{equation}

From Eq.(\ref{curren-cons1}) it follows that the norm is conservative only in the case,
\begin{equation}
j_1(0,t)=j_2(0,t)+j_3(0,t),
\label{curren-cons}
\end{equation}
which is the connection formula for the current density or the local current conservation condition at the vertex $O$. Similar conditions can be obtained for more complicated topologies. Below, Eq.(\ref{curren-cons}) will be evaluated explicitly.

Let us assume that a single (bright) soliton in PSG is described
with use of parts of ZSS (\cite{zakh}) lying on individual bonds
($k=1,2,3$) as follows:
\begin{equation}
\Psi_k(x,t)=\frac{a_k\sqrt 2}{\sqrt\beta_k}\cdot
\frac{\exp\left[i\frac{v_k}{2}x-i\left(\frac{v_k^2}{4}-a_k^2\right)t\right]}
{\cosh \left[a_k(x+l_k-v_kt)\right]},
\label{1-soliton}
\end{equation}
where $v_k, -l_k$ and $a_k$  are arbitrary parameters characterizing velocity, initial center of mass and amplitude of a soliton, respectively. While these solutions are finite at the origin, it tends to zero at the infinity.

We shall obtain the condition for the soliton solution Eq.(\ref{1-soliton}) to satisfy Eq.(\ref{curren-cons}).
Noting
\begin{equation}
\frac{\partial\Psi_k}{\partial
x}=\left[i\frac{v_k}{2}-a_k\tanh(a_k(x+l_k-v_kt))\right]\Psi_k,
\label{Ldx}
\end{equation}
the current density becomes
\begin{equation}
j_k(x,t)=v_k
|\Psi(x,t)|^2=\frac{2v_ka_k^2}{\beta_k\cosh^2\left[a_k(x+l_k-v_kt)\right]}.
\label{current-expl}
\end{equation}
Then Eq.(\ref{curren-cons}) is expressed as
\begin{equation}
\frac{v_1a_1^2}{\beta_1\cosh^2\left[a_1(l_1-v_1t)\right]}=
\frac{v_2a_2^2}{\beta_2\cosh^2\left[a_2(l_2-v_2t)\right]}+
\frac{v_3a_3^2}{\beta_3\cosh^2\left[a_3(l_3-v_3t)\right]}.
\label{curren-cons-f}
\end{equation}

Because of the linear independence of three functions, the connection formula Eq. (\ref{curren-cons-f}) is satisfied only in the following three cases:
\begin{equation}\label{eq7}
\mbox{(i)} \quad a_3v_3= 0,\quad a_1l_1=a_2l_2,\quad
a_1v_1=a_2v_2, \quad \frac{a_1}{\beta_1}=
\frac{a_2}{\beta_2},\qquad \
\end{equation}
\begin{equation}\label{eq8}
\mbox{(ii)} \quad a_2v_2= 0,\quad a_1l_1=a_3l_3,\quad
a_1v_1=a_3v_3, \quad \frac{a_1}{\beta_1}=
\frac{a_3}{\beta_3},\qquad \
\end{equation}
\begin{equation}\label{eq9}
\mbox{(iii)} \quad a_1l_1= a_2l_2=a_3l_3, \quad
a_1v_1=a_2v_2=a_3v_3, \quad
\frac{a_1}{\beta_1}=\frac{a_2}{\beta_2}+\frac{a_3}{\beta_3}.
\end{equation}
The above results are also available more straightforwardly by demanding the norm Eq.(\ref{norm}) to be time-independent.
The cases  (i) and (ii) suggest that a soliton simply moves from the bond $b_1$ to bonds $b_2$ and $b_3$, respectively, while the case  (iii) is indicative of the splitting of a soliton on $b_1$ into two parts with one appearing on $b_2$ and the other on $b_3$.
Besides  Eqs.(\ref{eq7})-(\ref{eq9}), there are additional constraints on solitonic parameters and strength of nonlinearity by noting the global current conservation,
which will be described below.

\subsection{Current Conservation Rule and Sum Rule for Strength of Nonlinearity}

We then demand the total current of the soliton in PSG, which can be calculated as
\begin{equation}
J=\int\limits_{-\infty}^0j_1(x,t)dx+
\int\limits_0^{+\infty}j_2(x,t)dx+\int\limits_0^{+\infty}j_3(x,t)dx
\nonumber\end{equation}
\begin{equation}
=\sum_{k=1}^3(-1)^{2^{k-1}+1}\frac{2v_ka_k}{\beta_k}\tanh(a_k(l_k-v_kt))+\sum_{k=1}^3\frac{2v_ka_k}{\beta_k}.
\label{tot-curr}
\end{equation}

It follows from Eq. (\ref{tot-curr}) that the
total current is conserved, if the additional conditions
\begin{equation}
\frac{v_1a_1}{\beta_1}=\frac{v_2a_2}{\beta_2},
\label{tot-curre1-2}
\end{equation}
\begin{equation}
\frac{v_1a_1}{\beta_1}=\frac{v_3a_3}{\beta_3},
\label{tot-curre1-3}
\end{equation}
and
\begin{equation}
\frac{v_1a_1}{\beta_1}=\frac{v_2a_2}{\beta_2}+\frac{v_3a_3}{\beta_3},
\label{tot-curre1}
\end{equation}
 are satisfied for cases (i), (ii) and (iii), respectively.

The norm conservation rule in Eqs.(\ref{eq7})-(\ref{eq9}) and current conservation rule
in Eqs.(\ref{tot-curre1-2})-(\ref{tot-curre1})
can be simultaneously satisfied under the following cases:
\begin{equation}\label{beta-2}
\mbox{[i]} \quad a_3= 0,\quad a_1=a_2=a; v_1=v_2=v;  l_1=l_2=l,
\quad \beta_1=\beta_2=\beta,
\end{equation}
\begin{equation}\label{beta-3}
\mbox{[ii]} \quad a_2= 0,\quad a_1=a_3=a; v_1=v_3=v;  l_1=l_3=l,
\quad \beta_1=\beta_3=\beta,
\end{equation}
\begin{equation}\label{beta}
\mbox{[iii]} \quad a_1=a_2=a_3=a; v_1=v_2=v_3=v;  l_1=l_2=l_3=l, \nonumber\end{equation}
\begin{equation}
\qquad \frac{1}{\beta_1}=\frac{1}{\beta_2}+\frac{1}{\beta_3},
\end{equation}
where $a,v,l$ and $\beta$ are arbitrary constants.
Equations (\ref{beta-2})-(\ref{beta}) imply: Firstly, the soliton solution at each bond should be a part of the identical ZSS, namely the main solitonic parameters should be common to individual bonds (line segments) in the graph  except for the strength of nonlinearity. In particular, solitons on $b_2$ and $b_3$ are initially located outside these bonds;  Secondly, the soliton can bifurcate in passing  through the vertex if the strengths of nonlinearity $\beta_j$ at individual bonds $b_j$ satisfy the sum rule in Eq.(\ref{beta})! This rule can also be obtained in a different way with use of symmetry argument in solving DNLSE (see Appendix).

To be explicit, we shall see the following dynamics:
In the first two cases, [i] and [ii], the soliton coming first from the bond
$b_1$ disappears or becomes a ghost at time $\tau \equiv \frac{l}{v}$, when a new soliton appears in either one of  $b_2$ and $b_3$. In these cases,
we have $\beta_1=\beta_2=\beta$ or $\beta_1=\beta_3=\beta$,  and therefore the soliton propagation is nothing but that in an ideal 1-d chain.

\begin{figure}[htb]
\centerline{\includegraphics[width=8cm]{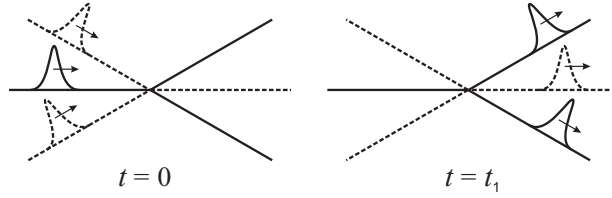}}
\caption{Splitting of Soliton. $t_1 > \tau$. Broken curves
represent ghost solitons.} \label{split}
\end{figure}

The third case  [iii] is the most interesting, where the  soliton at bond $b_1$ splits into two parts and
appears in both of  $b_2$ and $b_3$, as shown in Fig.\ref{split}. This is a novel feature of the soliton propagation through a branched chain and networks in general.
Precisely speaking, the soliton dynamics here is governed by a single characteristic time $\tau \equiv \frac{l}{v}$.
While for $0\le t \le \tau$ the soliton at $b_1$  is a real one and those at $b_2$ and $b_3$ are ghosts,  for $\tau \le t$ the soliton at $b_1$ is a ghost and those
at $b_2$ and $b_3$ are real. Common to the cases [i]-[iii], the incoming real soliton on $b_1$ and outgoing ghost solitons at $b_2$ and $b_3$ arrive at the vertex $O$
at time $\tau$ simultaneously.

In a similar way, the time-reversal process is obvious that two solitons
start to move from bonds $b_2$ and $b_3$, meet each other at the vertex at time $\tau$ and move as a single soliton
along the bond $b_1$ towards $-\infty$.

\subsection{Boundary Condition at Vertex}

In closing this Section,  we should clarify the nature of the boundary condition at the vertex.
Equations (\ref{beta-2})-(\ref{beta}) have led to the issue that the solution on each bond is a part of the universal soliton solution on a simple 1-d chain but multiplied by the inverse of square root of bond-dependent nonlinearity.
Indeed, under the condition in Eq. (\ref{beta}), the soliton
solution in Eq. (\ref{1-soliton}) can be written as
\begin{equation}
\Psi_k(x_k,t)=\sqrt{\frac{2}{\beta_k}}i q(x_k,t),
\label{scale-func}
\end{equation}
where the functions $ q(x_1,t)$ and $q(x_{2,3},t)$ are restricted to
$(-\infty ; 0]$ and
$[0,+\infty)$, respectively and stand for the individual part of the $\beta$-independent universal solution $q(x,t)$
which satisfies the nonlinear Schr\"odinger equation with $\beta=2$,
\begin{equation}
iq_t+q_{xx}+2q|q|^2=0, \qquad  -\infty<x<+\infty.
\label{betaless-eq}
\end{equation}
Here, $\Psi (x,t)$ itself is neither continuous  nor smooth at the
vertex. As for $q(x)$, however , there is no singularity there:
$\lim_{x_1 \to -0}q^{(n)}(x_1)=\lim_{x_{2,3} \to
+0}q^{(n)}(x_{2,3})$ for any $n=0,1,2, \cdots$ with $q^{(n)}$ the
$n$-th order derivative of $q$. Equivalently, the solution scaled
by by $\beta_k^{1/2}$ ($k=1,2,3$) satisfies the boundary condition
at the vertex:
\begin{equation}
\lim_{x_1 \to -0}\beta_1^{1/2}\Psi_1^{(n)}(x_1)=\lim_{x_{2,3} \to
+0}\beta_{2,3}^{1/2}\Psi_{2,3}^{(n)}(x_{2,3}),
\label{smooth1}\end{equation} for any $n=0,1,2, \cdots$ with
$\Psi_k^{(n)}$ the $n$-th order derivative of $\Psi_k$. The
absence of singularity of any kind in the scaled  function
$\beta_k^{1/2}\Psi_1^{(n)}(x_k)$ is the reason why we see
neither reflection nor emergence of radiation at the vertex, as is evidenced numerically in Fig.\ref{numerical} (see also Appendix).

This implies that the boundary conditions used in the present work are
different from those in the case of time-dependent linear
Sch\"odinger equation on graphs (for example see, Kottos \&
Smilansky 1997). If we would take ordinary continuity and
smoothness conditions, e.g.,
\begin{equation}
\Psi_k(0,t)=\Psi_l(0,t),
\nonumber\end{equation}
\begin{equation}
\frac{\partial\Psi_1}{\partial x}(0,t)=
\frac{\partial\Psi_2}{\partial x}(0,t)+
\frac{\partial\Psi_3}{\partial x}(0,t), \label{smooth2}
\end{equation}
we shall see a completely different nonlinear dynamics of solitons such as reflection of a soliton at the vertex. The initial value problem under the conditions in Eq.(\ref{smooth2}) at the vertex will be treated elsewhere.

\section{Energy and Other Conservation Rules}

We now proceed to the calculation of the energy of soliton in the graph. The total energy can be evaluated as
\begin{equation}
E=\int\limits_{-\infty}^0\left(
\left|\frac{\partial\Psi_1(x,t)}{\partial
x}\right|^2-\frac{\beta_1}{2}|\Psi_1(x,t)|^4 \right)dx+\nonumber\end{equation}
\begin{equation}
\int\limits_0^{+\infty}\left(
\left|\frac{\partial\Psi_2(x,t)}{\partial
x}\right|^2-\frac{\beta_2}{2}|\Psi_2(x,t)|^4 \right)dx+\nonumber\end{equation}
\begin{equation}
\int\limits_0^{+\infty}\left(
\left|\frac{\partial\Psi_3(x,t)}{\partial
x}\right|^2-\frac{\beta_3}{2}|\Psi_3(x,t)|^4 \right)dx.
\label{en1}
\end{equation}
With use of Eqs.(\ref{1-soliton}) and (\ref{Ldx}), we have
\begin{equation}
\left|\frac{\partial\Psi_k(x,t)}{\partial
x}\right|^2-\frac{\beta_k}{2}|\Psi_k(x,t)|^4=
\left(\frac{v_k^2}{4}-a_k^2\right)+\nonumber\end{equation}
\begin{equation}
2a_k^2\tanh^2[a_k(x+l_k-v_kt)]\cdot\frac{1}{\cosh^2[a_k(x+l_k-v_kt)]}.
\label{en2}
\end{equation}
In the most essential case [iii], for example,  Eqs. (\ref{en1}) and (\ref{en2}) together with Eq.(\ref{beta}) lead to:
\begin{equation}
E=2a\left(\frac{1}{\beta_1}-\frac{1}{\beta_2}-\frac{1}{\beta_3}\right)
\left[\left(\frac{v^2}{4}-a^2\right)\tanh(al-avt)+
\frac{2a^2}{3}\tanh^3(al-avt)\right]\nonumber\end{equation}
\begin{equation}
+\sum_{k=1}^{3}\frac{2a}{\beta_k}\left(\frac{v^2}{4}-\frac{a^2}{3}\right),
\label{ener-red}
\end{equation}
which proves constant under the sum rule in Eq.(\ref{beta}).

One can generalize the argument so far beyond a single-soliton
solution: So long as the general solution on PSG is described by
parts of the corresponding universal scaled function $q(x,t)$ as
shown in Eq.(\ref{scale-func}), all the conservation laws for 1-d
chain hold for PSG under the sum rule Eq.(\ref{beta}). Applying
Zakharov-Shabat's theorem (\cite{zakh}), we obtain the general
conservation rules
\begin{equation}
(2i)^nC_n\beta_1^{-1}=\sum\limits_{k=1}^{3}\beta_k^{-1}
\int_{b_k}f_n(q_k(x_k, t))dx_k,
\label{general-cons}
\end{equation}
where $C_n$ are constant, and $f_n(q(x))$ obeys the recursion
relation (see Eq.(35) of \cite{zakh}):
\begin{equation}
f_{n+1}=q\frac{\partial}{\partial
x}\left(\frac{1}{q}f_n\right)+\sum_{j+l=n}f_jf_l,\quad f_1=|q|^2.
\label{recursion-rel}
\end{equation}

In fact, with use of Eq. (\ref{beta}), the r.h.s. of Eq.(\ref{general-cons}) turns out:

$$\beta_1^{-1}\int\limits_{-\infty}^{0}
f_n(q(x,t))dx+(\beta_2^{-1}+
\beta_3^{-1})\int\limits_{0}^{+\infty} f_n(q(x,t))dx$$
\begin{equation} =
\beta_1^{-1}\int\limits_{-\infty}^{+\infty}
f_n(q(x,t))dx=\beta_1^{-1}(2i)^nC_n \quad , \label{proof}
\end{equation}
where the second equality comes from the conservation rule for the
1-d chain (\cite{zakh}).

It is easy to see that $f_n$ is the $2n$-th order polynomial of $q$
and its derivatives with respect to $x$, written in the
following form
\begin{equation}
f_n=\sum_{s=1}^{n}b_sP_{n,2s}(q, q_x, q_{xx},\cdots),
\label{polynom}
\end{equation}
where $P_{n,2s}=q^{k_1}(q^*)^{k_2}q_x^{k_3}(q_x^*)^{k_4}\cdots $
with $ k_1+k_2+k_3+\cdots=2s$.

In this way one can obtain an infinite number of conservation laws in PSG,
\begin{equation}
(2i)^nC_n\beta_1^{-1}=  frac{1}{2}\sum_{k=1}^{3}
\int\limits_{b_k}\sum_{s=1}^n
b_s\left(\frac{\beta_k}{2}\right)^{s-1} P_{n,2s}(\Psi_k,
\Psi_{k,x}, \cdots)dx_k. \label{infinite-cons}
\end{equation}

In Eq.(\ref{infinite-cons}), the cases $n=1,2$, and  $3$ give the norm, current and energy conservation rules in Eqs.(\ref{norm}), (\ref{tot-curr}) and (\ref{en1}), respectively. Some higher-order conservation rules are as follows:
\begin{equation}
(2i)^4 C_4\beta_1^{-1}=
\frac{1}{2}\sum_k\int\limits_{b_k}\left(\Psi_k\frac{\partial^3\Psi_k^*}{\partial
x_k^3}+\frac{3\beta_k}{2}\Psi_k \frac{\partial \Psi_k^*}{\partial
x_k}|\Psi_k|^2\right)(x_k,t)dx_k, \label{const-4}
\end{equation}
\begin{equation}
(2i)^5 C_5\beta_1^{-1}= \frac{1}{2}\sum_k\int\limits_{b_k}\left[
\left|\frac{\partial^2\Psi_k}{\partial x_k^2} \right|^2
+\frac{\beta_k^2}{2}|\Psi_k|^6-
\frac{\beta_k}{2}\left(\frac{\partial}{\partial x_k
}|\Psi_k|^2\right)^2\right. \nonumber\end{equation}
\begin{equation}
\left. -3\beta_k\left|\frac{\partial\Psi_k}{\partial
x_k}\right|^2|\Psi_k|^2 \right](x_k,t) dx_k. \label{const-5}
\end{equation}

The above treatment is also true for more general star graphs consisting of $N$ semi-infinite bonds connected at a single vertex. In such cases, the initial soliton at a bond splits into $N-1$ solitons in the remaining bonds, and the extended version of Eq. (\ref{beta}) is:
\begin{equation}\label{beta-g}
a_j=a, \quad v_j=v, \quad  l_j=l  \quad (j=1,2,\cdots,N),\ \ \
\frac{1}{\beta_1}=\sum_{j=1}^{N-1}\frac{1}{\beta_j}.
\end{equation}
where $a,v$ and $l$ are arbitrary constants.

All the soliton solutions by Zakharov and Shabat, as they stand, are applicable to networks, but, in order to see the bifurcation of the soliton solution at vertices, the strengths of nonlinearity at individual bonds should be different and satisfy the sum rule Eqs.(\ref{beta}) or (\ref{beta-g}) at each of vertex, which we shall see in details in the following Sections.

\section{Multi-Soliton Solutions}

For the case when the condition like Eq.(\ref{beta}) will be satisfied,
one can construct more general multi-soliton solutions of NLSE satisfying all conservation rules.
Below we shall give such multi-soliton solutions on PSG.

\subsection{Soliton Collision Described by Double-Soliton Solution}
Here, three solitons incoming along each of 3 bonds towards the vertex are shown to be scattered to three directions.
The following solution of NLSE on PSG describes double-soliton solution or a soliton
collision:
\begin{equation}
\Psi_1(x_1,t)=\sqrt {\frac{2}{\beta_1}}i q(x_1,t), \quad
x_1\in (-\infty, 0],\nonumber\end{equation}
\begin{equation}
\Psi_k(x_k,t)=\sqrt{\frac{2}{\beta_k}}i q(x_k,t), \quad x_k
\in[0,+\infty), \qquad k=1,2. \label{Nsol}
\end{equation}
Here
$$
q(x,t)=A_+(x,t)\exp(-iz_+^*x)+A_-(x,t)\exp(-iz_-^*x)
$$
where
$$
A_{\pm}=\frac{D_{\pm}B_{\mp \mp}-D_{\mp}B_{\pm
\mp}}{B_{++}B_{--}-B_{+-}B_{-+}}
$$
with
$$
D_{\pm}=-c_{\pm}^*\exp(-iz_{\pm}^*x),$$
$$
B_{\pm
\mp}=\sum_{l=+,-}\frac{c_{l}c_{\pm}^*}{(z_l-z_+^*)(z_l-z_-^*)}\exp(i(2z_l-z_+^*-z_-^*)),
$$
$$
B_{\pm
\pm}=\sum_{l=+,-}\frac{c_{l}c_{\pm}^*}{(z_l-z_{\pm}^*)^2}\exp(i(2z_l-2z_{\pm}^*))-1,
$$
$$
z_{\pm}=\xi_{\pm}+i\eta_{\pm}= const, $$
$$
 c_{\pm}=c_{\pm}(0)\exp(4iz_{\pm}^2t),
$$
where $\xi_{\pm}, \eta_{\pm}$ and $c_{\pm}(0) $ are scattering data characterizing the solitonic parameters.
\begin{figure}[htb]
\centerline{\includegraphics[width=8cm]{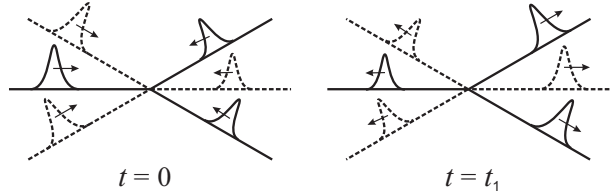}}
\caption{Collision of solitons. $t_1>\tau$. Broken curves
represent ghost solitons.} \label{dsoliton}
\end{figure}

A collision among solitons is possible for  $\xi_{+}\xi_{-}<0$.
It follows from the above double-soliton solution
that after a collision two solitons penetrate each other and  are split
into two independent solitons. Velocities of
solitons are characterized by the parameters $\xi_{\pm}$,
while its amplitude depends on the parameters, $\beta_k$,
and $\eta_{\pm}$. On the bond $b_1$, for example, the double solitons are regarded
as consisting from  a real one with velocity $v_+>0$ on the bond $b_1$ and  a ghost one with  velocity $v_-<0$ on the ghost bond
extended from $b_1$ to the positive $x$ region.
Let consider the case that a pair of real and ghost solitons meet at the vertex
at time $\tau$, by properly
choosing the initial position of the centers of mass which depends on
the scattering data.  Then we can see collision among three real solitons in Fig.\ref{dsoliton}:  While for $t<\tau$
the real soliton on $b_1$ ($b_{2,3}$) has the velocity $v=v_+$ ($v=v_-$), for $t>\tau$ it acquires $v=v_-$ ( $v=v_+$),
describing a collision among 3 real solitons.

\subsection{$N$-Soliton Solution.}
Similarly to the case of two-soliton problem we can obtain $N$-soliton solution of the NLSE on PSG by assuming
Eqs.(\ref{beta}) and (\ref{scale-func}).
Let $q(x,t)$ be the $N$-soliton solution of NLSE with the nonlinearity $\beta=2$ in Eq.(\ref{betaless-eq}).
Then the $N$-soliton solution of Eq.(\ref{eq1}) on PSG can be
constructed by multiplying the corresponding universal solution $g(x,t)$ with the bond-dependent factor $\sqrt{\frac{2}{\beta_k}}i$ on individual bonds $b_k$.

\begin{figure}[htb]
\centerline{\includegraphics[width=85mm]{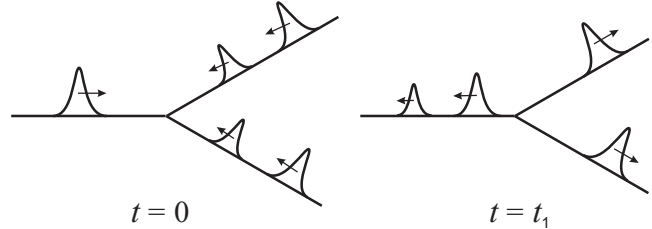}}
\caption{Collision of multi-solitons, which apparently breaks
conservation of particle numbers. $t_1>\tau$. $N=3, M=2$. }
\label{multi-soliton}
\end{figure}

It should be noted that at arbitrary  moment of time the bonds $b_2$ and $b_3$  have the same number of solitons, $M$.
However, at the same time the bond $b_1$ has  $N-M$ solitons, as exemplified in Fig.\ref{multi-soliton}.
With use of $N$-soliton solutions, a variety of splitting of solitons at the vertex
is found depending on the initial velocities, which apparently breaks the conservation of particle numbers.
As already described in Section IV, all the conservation laws hold for this solution.

\section{Other Types of Graphs}
Now we proceed to explore soliton solutions of NLSE on other kind of graphs and explore the sum rule and conservation rules
for solitons to propagate through these graphs.

\begin{figure}[htb]
\centerline{\includegraphics[width=75mm,
height=45mm]{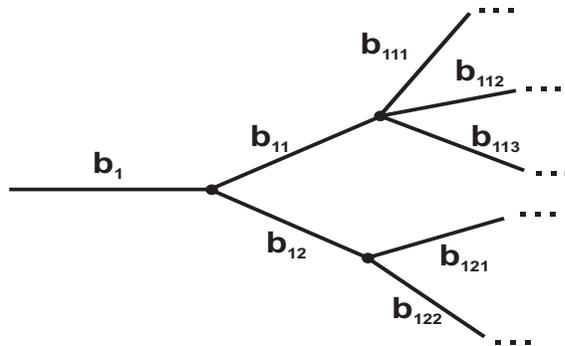}} \caption{Tree graph. $b_1\sim
(-\infty, 0), b_{11}\sim (0, L_{11}), b_{12} \sim (0,L_{12})$, and
$b_{1ij} \sim (0,+\infty)$ with $i,j=1,2, \cdots$.} \label{tree-0}
\end{figure}

An example of the graph for which the soliton solution of NLSE can be
obtained analytically is a tree graph  in
Fig. \ref{tree-0}.  Hereafter, for an arbitrary one of bonds in the tree graph,  we shall employ
an abbreviation like $b_\Gamma\equiv b_{1ij \cdots}$.
On each bond $b_\Gamma$ we have NLSE given by Eq.
(\ref{eq1}) and for each vertex the following conditions is
satisfied:
\begin{equation}
\frac{1}{\beta_\Gamma}=\sum_k\frac{1}{\beta_{\Gamma k}},
\label{sum-tree}
\end{equation}
which is again available from the norm and current conservation rules.
The soliton solution satisfying these conditions can be written as
\begin{equation}
\Psi_{\Gamma}(x_{\Gamma},t)=\sqrt{\frac{2}{\beta_{\Gamma}}}iq(x_{\Gamma}+ s_{\Gamma},t; s_{\Gamma}), \qquad x_{\Gamma}\in b_{\Gamma}.
\label{tree}
\end{equation}


Here paremeter $s_{\Gamma}$ is the length of the path that soliton
passes from $b_1$ through $b_{\Gamma}$. For tree graphs this
parameter is given as $$ s_1=s_{1i}=l,\quad s_{1ij}=l+L_{1i},$$

\begin{equation} s_{\Gamma}\equiv s_{1ij \cdots
lm}=l+L_{1i}+L_{1ij}+\cdots+ L_{1ij \cdots l}, \label{cayley}
\end{equation}
where $-s_{\Gamma}$ represents an initial location of the solution (arbitrary part), and
$L_{1i}, L_{1ij}$, $\cdots ,L_{1ij \cdots l}$ are lengths of the finite bonds prior to $b_{\Gamma} \equiv b_{1ij \cdots lm}$.

Below, applying the induction method, we give a proof of conservation rules  for soliton solutions of NLSE on any tree graph. Let us denote the tree graph in Fig.\ref{tree-0} as $G$ and assume the conservation rules to hold in $G$:  ${\sum}_{b_{\Gamma} \in G} \beta_{\Gamma}^{-1}\int\limits_{b_{\Gamma}} f_n(q(x_{\Gamma}+s_{\Gamma},t))dx_{\Gamma}=(2i)^nC_n\beta_1^{-1}$.  Then we construct an enlarged tree graph in the following way:  First, cut an arbitrary one of the right-most semi-infinite chain $b_{\Lambda}\sim (0, +\infty)$ at a point $A$ located by distance $L_{\Lambda}$ from the nearest vertex and then attach $N$ semi-infinite bonds to the point $A$ which now becomes a new vertex point. Namely the bond $b_{\Lambda}$ is now replaced by the finite bond  $\hat{b}_{\Lambda} \sim (0,L_{\Lambda})$ connected with $N$ semi-infinite bonds $\hat{b}_{{\Lambda} m} \sim (0,+\infty)$ with $m=1,\cdots, N$.  The enlarged tree graph thus obtained is denoted as $G'$. In the same way as in Eq.(\ref{general-cons}), the general conserved quantity for $G'$ is given by
$$
{\sum}_{b_{\Gamma} \in G-b_{\Lambda}}
\beta_{\Gamma}^{-1}\int\limits_{b_{\Gamma}}
f_n(q(x_{\Gamma}+s_{\Gamma},t))dx_{\Gamma}
+\beta_{\Lambda}^{-1}\int\limits_{\hat{b}_{\Lambda}}
f_n(q(x_{\Lambda}+s_{\Lambda},t))dx_{\Lambda}
$$
$$ +\sum_{m=1}^{N}\beta_{{\Lambda}
m}^{-1}\int\limits_{\hat{b}_{{\Lambda} m}} f_n(q(x_{{\Lambda}
m}+s_{{\Lambda} m}+L_{{\Lambda} m},t))dx_{{\Lambda} m}
$$
$$
={\sum}_{b_{\Gamma} \in G}
\beta_{\Gamma}^{-1}\int\limits_{b_{\Gamma}}
f_n(q(x_{\Gamma}+s_{\Gamma},t))dx_{\Gamma}
$$
 $$
-\beta_{\Lambda}^{-1}\int\limits_{L_{\Lambda}}^{+\infty}f_n(q(x+s_{\Lambda},t))dx
+\sum_{m=1}^{N}\beta_{{\Lambda}
m}^{-1}\int\limits_{L_{\Lambda}}^{+\infty}f_n(q(x+s_{\Lambda},t))dx
$$

$$ =(2i)^nC_n\beta_1^{-1}-
\left(\beta_{\Lambda}^{-1}-\sum_{m=1}^N\beta_{{\Lambda}
m}^{-1}\right)\int\limits_{L_{\Lambda}}^{+\infty}f_n(q(x+s_{\Lambda},t))dx.
$$
Here  $\sum_{b_{\Gamma} \in G-b_{\Lambda}}$ and $\sum_{b_{\Gamma} \in G}$ imply summations over all bonds in
$G$ except for $b_{\Lambda}$ and over all bonds of $G$, respectively.
It is clear that the final expression becomes constant $(2i)^nC_n\beta_1^{-1}$ under the sum rule in Eq.(\ref{sum-tree}).
Thus, starting from PSG in Fig. \ref{star} and repeating the above procedure,
we can get the conservation rules for all  possible tree graphs.

\begin{figure}[htb]
\centerline{\includegraphics[width=85mm]{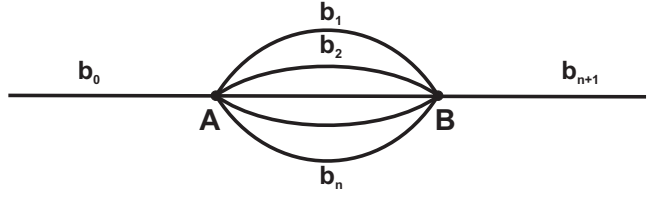}} \caption{Graph
with loops. $b_0\sim (-\infty, 0), b_{n+1}\sim (0, +\infty), b_k
\sim (0,L)$ with $k=1,2,...,n$.} \label{loop}
\end{figure}

Another example for which soliton can be easily obtained is a graph
with loops (see Fig.\ref{loop}). This graph consists of two
semi-infinite bonds whose edges are connected with $n$ bonds having
finite lengths. Again, requiring the following conditions for the
coefficients of NLSE:
$$\frac{1}{\beta_0}=\sum_{k=1}^{n}\frac{1}{\beta_k}=\frac{1}{\beta_{n+1}}$$
we can write the soliton solution by Eqs. (\ref{tree}).

\begin{figure}[htb]
\centerline{\includegraphics[width=85mm]{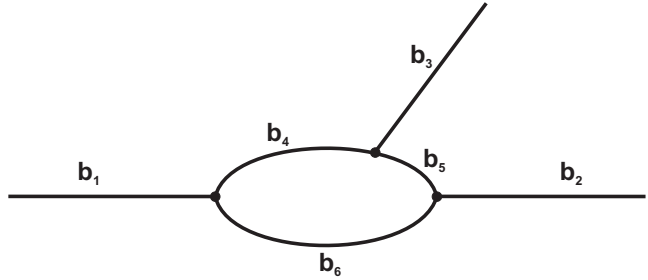}}
\caption{Loop with semi-infinite bonds. $b_1\sim (-\infty, 0),
b_{2}, b_3\sim (0, +\infty), b_k\sim (0,L_k)$ with $k=4,5,6.$
$L_6=L_4+L_5$.} \label{circle}
\end{figure}

Also, the exact soliton solution can be obtained for the graph in Fig.\ref{circle} where the corresponding condition  for
the parameters, $\beta_k$ is required. This graph can be considered as a loop graph connected with 3 semi-infinite bonds.

\begin{figure}[htb]
\centerline{\includegraphics[width=80mm,
height=40mm]{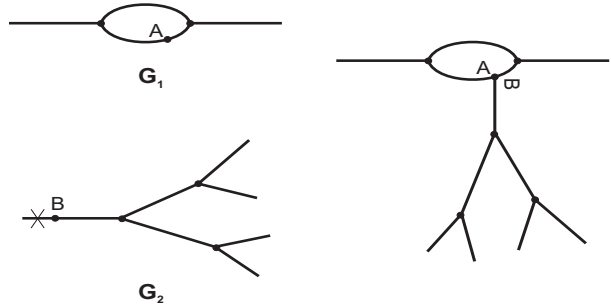}} \caption{Combination (right
panel) of loop $G_1$ and tree $G_2$ graphs (left panel).}
\label{comb1}
\end{figure}

Combining the above topologies one can construct different graphs(having semi-infinite bonds) for which soliton solution of NLSE can be constructed.
To do this consider two graphs
$G_1$ and $G_2$ where there exist soliton solutions given by Eq.(\ref{tree}). Soliton solution on the combination of the graphs $G_1$ and $G_2$
can be constructed using one of two methods below.

Method I. Let  $A$ be a point on a finite bond of the
graph $G_1$ and  $B$ be a point on a semi-infinite bond of $G_2$.
Connecting two graphs by putting together the
points $A$ and $B$, we reach the graph in Fig. \ref{comb1}.
\begin{figure}[htb]
\centerline{\includegraphics[width=85mm,
height=45mm]{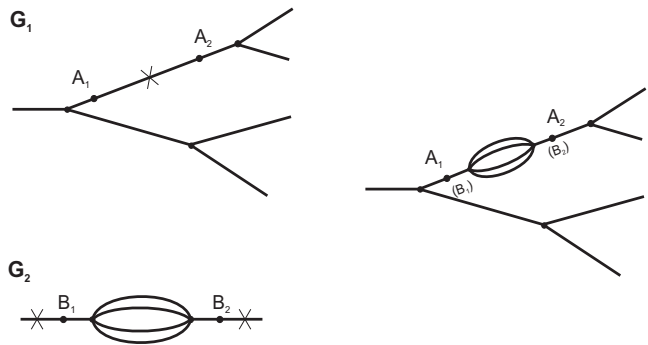}} \caption{Same as Fig.\ref{comb1},
but a different combination.} \label{comb2}
\end{figure}

Method II. Fixing two points $A_1$ and $A_2$  on a bond of graph
$G_1$  and cutting the part between these points and doing the
same thing with the points $B_1$ and $B_2$ of bond in the graphs
$G_2$ we can connect two graphs by putting together the points
$A_1$ , $B_1$ and $A_2$ and $B_2$, respectively, and thereby can reach a
new graph in Fig. \ref{comb2}.

In these different types of graphs, the soliton solution of NLSE can be constructed under the conditions given by Eqs.(\ref{sum-tree}) and (\ref{tree}). It should be noted that throughout in our approach the graphs are supposed to  have at least two semi-infinite bonds.

\section{Summary and discussions}
We have explored soliton solutions of nonlinear Schr\"odinger
equation (NLSE) on simple networks.  We first concentrated on
reflectionless propagation of a Zakharov-Shabat's soliton through
a branched chain, namely, a primary star graph. To satisfy the
conservation of the norm and global current  in the graph, the
solution on each bond should be a part of the universal soliton
solution on a simple 1-d chain but multiplied by the inverse of
square root of bond-dependent nonlinearity. Besides this,
nonlinearities at individual bonds around each vertex must satisfy
a sum rule: the inverse nonlinearity at an incoming bond should be
equal to the sum of inverse nonlinearities at the remaining bonds.
Under these conditions, all other conservation rules for solitons
in a simple 1-d chain have proved to hold for solitons propagating
through graphs. With use of Zakharov-Shabat's two-soliton
solutions we also find a collision among three solitons at the
vertex, and with use of $N$-soliton solutions, a variety of
splitting of solitons at the vertex was found depending on the
initial velocities, which apparently breaks the conservation of
particle numbers. The argument is extended to general star graphs,
tree graphs, loop graphs and their combinations. To see all
conservation rules to hold, a set of  inverse nonlinearities
should satisfy the generalized sum rule at each of vertices, which
we proved by the induction method.

So long as the sum rule for strength of nonlinearity at each
vertex holds, the boundary conditions (: connection formulas)
there for scaled wave functions are normal, and there is no
singularity at vertices that generates reflection and radiation.
Although there exist accumulation of analytical studies on initial
value problems on the semi-infinite chain (Ablowitz \& Segur 1975;
Fokas \textit{et al.} 2005) and the finite chain (Ramos \&
Villatoro 1994; Fokas \& Its 2004), no corresponding ones in
networks or graphs have appeared up to now. Under the boundary
connections different from the present paper, soliton dynamics
would become more complicated, whose analysis is a next
challenging subject.

\section*{Acknowledgments} We are grateful to F. Abdullaev, B.
Baizakov, M. Lakshmanan and E. Tsoy  for useful comments. The work
is partly supported through a project of the Uzbek Academy of
Sciences (FA-F2-084).

\section*{Appendix. Numerical method to solve nonlinear \\ Schr\"odinger equation
on primary star graph}
With use of space discretization ($\displaystyle \frac{\partial^{2}\psi}{\partial x^{2}}\ \Rightarrow\ \psi_{i+1}-2\psi_{i}+\psi_{i-1}$), the nonlinear Schr\"odinger equation in the 1-$d$ continuum with neither branches nor vertex
$$
 i\displaystyle \frac{\partial\psi}{\partial t}=-\frac{\partial^{2}\psi}{\partial
 x^{2}}-\beta|\psi|^{2}\psi
$$
can be reduced to
\begin{equation}
i\displaystyle \frac{\partial\psi_{i}}{\partial t}=-\psi_{i-1}+(2-\beta|\psi_{i}|^{2})\psi_{i}-\psi_{i+1},
\label{dnse-0}
\end{equation}
which is rewritten  in a matrix form as
\begin{equation}
i \frac{\partial}{\partial t} \left( \begin{array}{c}
\vdots\\
\psi_{i-1}\\
\psi_{i}\\
\psi_{i+1}\\
\vdots
  \end{array}\right)
  =H(t)
\left( \begin{array}{c}
\vdots\\
\psi_{i-1}\\
\psi_{i}\\
\psi_{i+1}\\
\vdots
\end{array}\right)
 \label{dns1}
 \end{equation}
with
\begin{equation}
 H(t)\equiv
\left( \begin{array}{ccccccc}
 &\ &(i-1) &(i) &(i+1) &\ & \\
 &\  &\vdots &\vdots &\vdots &\ & \\
\cdots&-1 &2-\beta|\psi_{i-1}|^{2} &-1 & 0&0&\cdots  \\
\cdots&0 &-1 & 2-\beta|\psi_{i}|^{2}& -1&0 &\cdots \\
\cdots&0 &0&-1 &2-\beta|\psi_{i+1}|^{2} &-1  &\cdots \\
& &\vdots  &\vdots  &\vdots  & &
\end{array}\right)
      \label{matrix-0}
\end{equation}

Then, by carrying out the time discretization with time difference $\Delta t$, Eq.(\ref{dns1}) reduces to
\begin{equation}
\left( \begin{array}{c}
\vdots\\
\psi_{i-1}(t+\Delta t)\\
\psi_{i}(t+\Delta t)\\
\psi_{i+1}(t+\Delta t)\\
\vdots
\end{array}\right)
  =\exp(-iH(t)\cdot\Delta t)
\left( \begin{array}{c}
\vdots\\
\psi_{i-1}(t)\\
\psi_{i}(t)\\
\psi_{i+1}(t)\\
\vdots
\end{array}\right)
 \label{dns2}
 \end{equation}
 It is obvious that Eq.(\ref{dns2}) conserves  the norm because of the
  unitarity of $\exp(-iH(t)\cdot\Delta t)$. It is sraight-forward that the diagonalization
 \begin{equation}
P^{-1}H(t)P= \left( \begin{array}{cccc}
\epsilon_1&0&\cdots&\cdots\\
0&\epsilon_2&0&\cdots\\
\vdots&\vdots& &
\end{array}\right)
 \label{dia-ener}
 \end{equation}
 gives rise to
 \begin{equation}
\exp(-iH(t)\cdot\Delta t)= \nonumber\end{equation}
\begin{equation}
P \left( \begin{array}{cccc}
\exp(-i\epsilon_1\Delta t)&0&\cdots&\cdots\\
0&\exp(-i\epsilon_2\Delta t)&0&\cdots\\
\vdots&\vdots& &  \end{array}\right)
 P^{-1}
 \label{dia-exp}
  \end{equation}

  In case of a branched chain, i.e., a primary star graph (PSG), we consider
  its discretized counterpart and introduce the numbering as in Fig.\ref{star}. Equations to generalize Eq.(\ref{dnse-0})  are given by
\begin{equation}
 i  \frac{\partial\psi_{i}^{(1)}}{\partial t}=-\psi_{i-1}^{(1)}+(2-\beta_{1}|\psi_{i}^{(1)}|^{2})\psi_{i}^{(1)}-\psi_{i+1}^{(1)},
\label{BB1}
\end{equation}
\begin{equation}
 i  \frac{\partial\psi_{j}^{(2)}}{\partial t}=-\psi_{j-1}^{(2)}+(2-\beta_{2}|\psi_{j}^{(2)}|^{2})\psi_{j}^{(2)}-\psi_{j+1}^{(2)},
  \label{BB2}
\end{equation}
\begin{equation}
i  \frac{\partial\psi_{k}^{(3)}}{\partial t}=-\psi_{k-1}^{(3)}+(2-\beta_{3}|\psi_{k}^{(3)}|^{2})\psi_{k}^{(3)}-\psi_{k+1}^{(3)},
 \label{BB3}
\end{equation}
which correspond to bonds $b_1$, $b_2$ and $b_3$, respectively.

The important problem is to search for the connection formula at the vertex, which will be resolved as follows.
Let call the end of $b_1$ as $K$ site. Similarly the starts of $b_2$ and $b_3$ are taken as
$L$ and $M$ sites, respectively. Introducing virtual wave functions $\psi_{L-1}^{(2)}$ and
$\psi_{M-1}^{(3)}$ and establish for their relationship with $\psi_{K}^{(1)}$.
As a manifold of the global solution on PSG, we assume
a discretized version of Eq.(\ref{scale-func}):
\begin{equation}
\psi_{j}(t)=\frac{\sqrt{2}}{\sqrt{\beta_k}}ig(x_j,t),
\end{equation}
where $k(=1,2,3)$ denotes individual bonds and the discrete lattice variable $i$ runs over PSG in Fig.\ref{star}. Because of the continuity of $g(x_j,t)$ at the vertex,  we obtain a connection formula:
\begin{equation}
\sqrt{\beta_{1}}\psi_{K}^{(1)}=\sqrt{\beta_{2}}\psi_{L-1}^{(2)}=\sqrt{\beta_{3}}\psi_{M-1}^{(3)}.
\end{equation}
On the other hand, with use of suitable parameters $s_{2}$ and $s_{3}$,
a virtual wave function  $\psi_{K+1}^{(1)}$ should be
\begin{equation}
 \psi_{K+1}^{(1)}=\frac{1}{s_{2}+s_{3}}\left(s_{2}\sqrt{\frac{\beta_{2}}{\beta_{1}}}\psi_{L}^{(2)}+s_{3}\sqrt{\frac{\beta_{3}}{\beta_{1}}}\psi_{M}^{(3)}\right).
 \end{equation}
Then Eqs.(\ref{BB1})-(\ref{BB3}) at the vertex  can be explicitly rewritten as
\begin{equation}
i \frac{\partial\psi_{K}^{(1)}}{\partial t}=-\psi_{K-1}^{(1)}+(2-\beta_{1}|\psi_{K}^{(1)}|^{2})\psi_{K}^{(1)} \nonumber\\
-\frac{1}{s_{2}+s_{3}}\left(s_{2}\sqrt{\frac{\beta_{2}}{\beta_{1}}}\psi_{L}^{(2)}+s_{3}\sqrt{\frac{\beta_{3}}{\beta_{1}}}\psi_{M}^{(3)}\right),
\label{KK}
\end{equation}
\begin{equation}
i  \frac{\partial\psi_{M}^{(2)}}{\partial t}=-\sqrt{\frac{\beta_{1}}{\beta_{2}}}\psi_{K}^{(1)}+(2-\beta_{2}|\psi_{M}^{(2)}|^{2})\psi_{M}^{(2)}-\psi_{M+1}^{(2)},
\label{MM}
\end{equation}
\begin{equation}
i \frac{\partial\psi_{L}^{(3)}}{\partial t}=-\sqrt{\frac{\beta_{1}}{\beta_{3}}}\psi_{K}^{(1)}+(2-\beta_{3}|\psi_{L}^{(3)}|^{2})\psi_{L}^{(3)}-\psi_{L+1}^{(3)}.
\label{LL}
\end{equation}

Lining up $\psi_{i}^{(1)}$, $\psi_{j}^{(2)}$ and $\psi_{k}^{(3)}$ vertically and rewriting Eqs.(\ref{BB1})-(\ref{BB3}) with Eqs.(\ref{KK})- (\ref{LL})
in a matrix form, we obtain the equation like Eq.(\ref{dns1}) with Eq.(\ref{matrix-0}),
but with a modified real matrix $\hat{H}(t)$. To conserve the norm, $\hat{H}(t)$ should be  symmetric, which imposes the following relationship:
\begin{equation}
 \frac{s_{2}}{s_{2}+s_{3}}\sqrt{\frac{\beta_{2}}{\beta_{1}}}=\sqrt{\frac{\beta_{1}}{\beta_{2}}},
\quad
\frac{s_{3}}{s_{2}+s_{3}}\sqrt{\frac{\beta_{3}}{\beta_{1}}}=\sqrt{\frac{\beta_{1}}{\beta_{3}}}.
\label{s2-s3}
\end{equation}
As a result, we have the sum rule for three kind of strengh of nonlinearity:
\begin{equation}
 \frac{1}{\beta_{1}}=\frac{1}{\beta_{2}}+\frac{1}{\beta_{3}},
 \label{beta-append}
 \end{equation}
which agrees with Eq.(\ref{beta}) obtained from the norm and current conservation rules for the PSG in the text.
Using Eq.(\ref{s2-s3}) , Eq.(\ref{KK}) can be replaced by
\begin{equation}
i  \frac{\partial\psi_{K}^{(1)}}{\partial t}=-\psi_{K-1}^{(1)}+(2-\beta_{1}|\psi_{K}^{(1)}|^{2})\psi_{K}^{(1)} \nonumber\\
-\left(\sqrt{\frac{\beta_{1}}{\beta_{2}}}\psi_{L}^{(2)}+\sqrt{\frac{\beta_{1}}{\beta_{3}}}\psi_{M}^{(3)}\right).
\label{KKK}
\end{equation}
By numerically solving Eqs.(\ref{BB1})-(\ref{BB3}) with Eqs.(\ref{KKK}), (\ref{MM}) and (\ref{LL}) under any initial
 condition, one obtains nonlinear dynamics of solitons without reflection at the vertex. Figure \ref{numerical} is obtained under the initial profile
 in Eq.(\ref{1-soliton}) with Eqs.(\ref{beta-append}) or (\ref{beta}).



\begin{thebibliography}{99}
\bibitem{har} F. Harary, {\it Graph Theory} (Addison-Wesley, Reading, 1969).
\bibitem{kott} T. Kottos and U. Smilansky, Phys. Rev. Lett. {\bf 79}, 4794 (1997); Ann. Phys. (NY) {\bf 274}, 76 (1999).
\bibitem{kiv} Y.S. Kivshar and G.P. Agarwal, {\it Optical Solitons: from Fibers to Photonic Crystals} (Academic Press, San Diego, 2003).
\bibitem{tro} A. Trombettoni and A. Smerzi, Phys. Rev. Lett. {\bf 86}, 2353 (2001).
\bibitem{binder} P. Binder {\it et al.,}  Phys. Rev. Lett. {\bf  84}  745 (2000).
\bibitem{bir} R. Burioni {\it et al.,}  Europhys. Lett. {\bf  52}  251 (2000).
\bibitem{yak} L.V. Yakushevich, A.V. Savin and L.I. Manevitch  Phys. Rev. E{\bf  66}, 016614 (2002).
\bibitem{fla} S. Flach and C. R. Willis, Phys. Rep. {\bf 295}, 181 (1998).
\bibitem{abl} M. J. Ablowitz, B. Prinari, and A. D. Trubatch, {\it Discrete and Continuous Nonlinear Schr\"odinger Systems} (University Press, Cambridge, 2004).
\bibitem{bur} R. Burioni, D. Cassi, P. Sodano, A. Trombettoni and A. Vezzani, Chaos {\bf 15}, 043501 (2005); Physica D {\bf 216}, 71 (2006).
\bibitem{pik} A.S. Pikovsky and D. L. Shepelyansky, Phys. Rev. Lett. {\bf 100}, 094101 (2008).
\bibitem{fla2} G. Kopidakis , S. Komineas , S. Flach and S. Aubry, Phys. Rev. Lett. {\bf 100}, 084103 (2008).
\bibitem{sulem} C. Sulem and P.L. Sulem, {\it The Nonlinear Schr\"odinger Equation} (Springer, New York, 1999).
\bibitem{zakh} V.B. Zakharov, A.B. Shabat, Sov. Phys. JETP.  {\bf 34} 62 (1972).


\bibitem{ablo} M.J.Ablowitz and H.Segur, J. Math. Phys. {\bf 16}, 1054 (1975).
\bibitem{foka1} A.S.Fokas, A.R.Its and L-Y Sung, Nonlinearity  {\bf 18},1771 (2005).
\bibitem{ramos} J.I.Ramos  and F.R.Villatoro, Mathl. Comput. Modelling.   {\bf 20}, 31 (1994).
\bibitem{foka2} A.S.Fokas and  A.R.Its, J. Phys. A: Math. Gen. {\bf 37}, 6091 (2004).



\end{thebibliography}
\end{document}